\documentclass[twocolumn,showpacs,preprintnumbers]{revtex4}
\usepackage{graphicx}% Include figure files
\usepackage{dcolumn}% Align table columns on decimal point
\usepackage{bm}% bold math
\begin{document}

%\begin{frontmatter}

\title{Soft X-ray Resonant Magnetic Scattering Studies
on Fe/CoO Exchange Bias System}
\author{Florin Radu, Alexei Nefedov,
Johannes Grabis, Gregor Nowak, Andre Bergmann, and  Hartmut Zabel}
\affiliation{Experimentalphysik/Festk\"orperphysik,
Ruhr-Universit\"at Bochum, D- 44780 Bochum, Germany}

\begin{abstract}

We have used soft X-ray Resonant Magnetic Scattering (XRMS) to
search for the presence of an effective ferromagnetic moment
belonging to the  antiferromagnetic (AF) layer which is in close
contact with a ferromagnetic (F) layer. Taking advantage of the
element specificity  of the XRMS technique, we have measured
hysteresis loops of both Fe and CoO layers of a
CoO(40~\AA)/Fe(150~\AA) exchange bias bilayer. From these
measurements we have concluded that the proximity of the F layer
induces a magnetic moment in the AF layer. The F moment of the AF
layer has two components: one  is frozen and does not follow the
applied magnetic field and the other one follows in phase the
ferromagnetic magnetization  of the F layer. The temperature
dependence of the F components belonging to the AF layer is shown
and discussed.

\end{abstract}

%\begin{keyword}
%\PACS 75.25+z\sep 75.60.Jk\sep 75.70.Cn\sep 61.10.Kw
 %\KEY  Exchange bias\sep X-ray resonant magnetic scattering \sep
%\end{keyword}
%\end{frontmatter}
 \maketitle
\section{Introduction}
Exchange bias refers to a shift of the ferromagnetic (F)
hysteresis loop to positive or negative values when a F system is
in contact with an antiferromagnetic (AF) system and cooled in an
applied magnetic field through the N\'eel temperature of the AF
system. The exchange bias phenomenon is associated with the
interfacial exchange coupling between ferromagnetic and
antiferromagnetic spin structures,  resulting in a unidirectional
magnetic anisotropy~\cite{bean:1956}. While the unidirectional
anisotropy was successfully introduced by Meiklejohn and
Bean(M\&B), the origin of the enhanced coercive field is yet not
well understood. The exchange bias effect is essential for the
development of magneto-electronic switching devices (spin-valves)
and for random access magnetic storage units. For these
applications a predictable, robust, and tunable exchange bias
effect is required.

Extensive data have been collected on the exchange bias field
$H_{EB}$ and the coercivity field $H_c$, for a large number of
bilayer systems, which are reviewed in Ref.
~\cite{berkowitz:1999,nogues:1999,stamps:2000,kiwi:2001}. The
details of the EB effect depend crucially on the AF/F combination
chosen and on the structure and thickness of the films. However,
some characteristic features apply to most systems: (1) $H_{EB}$
and $H_c$ increase as the system is cooled in an applied magnetic
field below the blocking temperature $T_B \leq T_N$ of the AF
layer, where $T_N$ is the N\'{e}el temperature of the AF layer;
(2) the magnetization reversal can be different for the ascending
and descending part of the hysteresis loop
\cite{radu:2002:1,gierlings:2002,lee:2002,radu:2003:1}, as was
first pointed out in reference \cite{fitzsimmons:2000}; (3)
thermal relaxation effects of $H_{EB}$ and $H_c$ indicate that a
stable magnetic state is reached only at very low temperatures
\cite{geoghegan:1998,goodman:2000,radu:2002:2}.

Several theoretical models have been developed for describing
possible mechanisms of the EB effect, including domain formation
in the AF layer with domain walls perpendicular to the AF/F
interface \cite{malozemoff:1987}, creation of uncompensated excess
AF spins at the interface \cite{schulthess:1998}, or the formation
of domain walls in the AF layer parallel to the interface
\cite{mauri:1987,kim:2005}. Another approach is the consideration
of diluted antiferromagnets in an exchange field. In the work of
Milt\'{e}nyi et al., Keller et al., and Nowak et
al.\cite{milt:2000,keller:2002,nowak:2002} the discussion about
compensated versus uncompensated interfacial spins is replaced by
a discussion of net magnetic moments within the antiferromagnetic
layer. The AF domains will carry a resulting magnetization which
will decrease non-exponentially with very high relaxation times.
This induced magnetization   in the AF layer is frozen and  within
the Domain State model it is responsible for the hysteresis shift.
Ohldag et al.~\cite{ohldag:2003}  and Kappenberger~et~al.~
\cite{Kappenberger:2003} observed  that  a small fraction of the
AF spins are uncompensated and responsible for the EB shift, as
predicted by M\&B model~\cite{bean:1956}. We concentrate on the
observation of  rotatable AF spins which contribute to the
enhanced coercivity, reported for almost all system described
in´the literature. They are essential for understanding the
coercivity enhancement as shown in Ref.~\cite{radu:2005:2}. There,
the coercivity was modelled by an extended M\&B model and assuming
an interface AF layer with variably anisotropy.

In this paper, hysteresis loops of the induced ferromagnetic
components belonging to the AF layer at the AF/F are described and
compared with the magnetization curves of the ferromagnet itself.

\section{Sample growth and hard x-ray characterization}

As substrate we used epi-polished single-crystalline
$(11\bar{2}0)$-oriented sapphire wafer. Before deposition, the
substrates were ultrasonically cleaned in acetone and ethanol and
then transferred into a high-vacuum rf-sputtering chamber which
provides  a base pressure of $1\cdot10^{-7}$~mbar. Prior to the
deposition the substrate were annealed at $500^{\circ}$~C for 3 h
and etched for 10 min with an Ar ion beam. Then the substrate was
cooled  down to $200^{\circ}$~C, where the Fe layer has been
deposited by rf-sputtering in Ar with a partial pressure of
$5\cdot10^{-3}$~mbar. Subsequently, the polycrystalline CoO layer
has been grown by sputtering of Co atoms in a mixture of Ar (94\%)
and O$_2$ (6\%). The deposition rates were 0.48 \AA/s and 0.57
\AA/s for Fe and CoO, respectively. The expected nominal
thicknesses, as calculated from deposition rates, were 150 \AA~
and 40 \AA~ for Fe and CoO, respectively. After deposition, the
structural quality of the bilayers was probed  by x-ray
diffraction and reflectivity, which is described further below.

X-ray scattering is most suitable for detailed structural
characterization
 of thin films and heterostructures. Information about the electron density profile perpendicular
 to the film plane is obtained
 via reflectivity measurements.
  High-angle radial scans at a
 reciprocal-lattice point
 (Bragg scans) provide information about the
 crystalline properties of the films.
The hard x-ray measurements were carried out with the use of
synchrotron radiation at the W1.1 beamline at HASYLAB. The
radiation wavelength was $\lambda$=1.5408 \AA.

\begin{figure}[!th]
\begin{center}
\includegraphics[clip=true,keepaspectratio=true,width=1\linewidth]{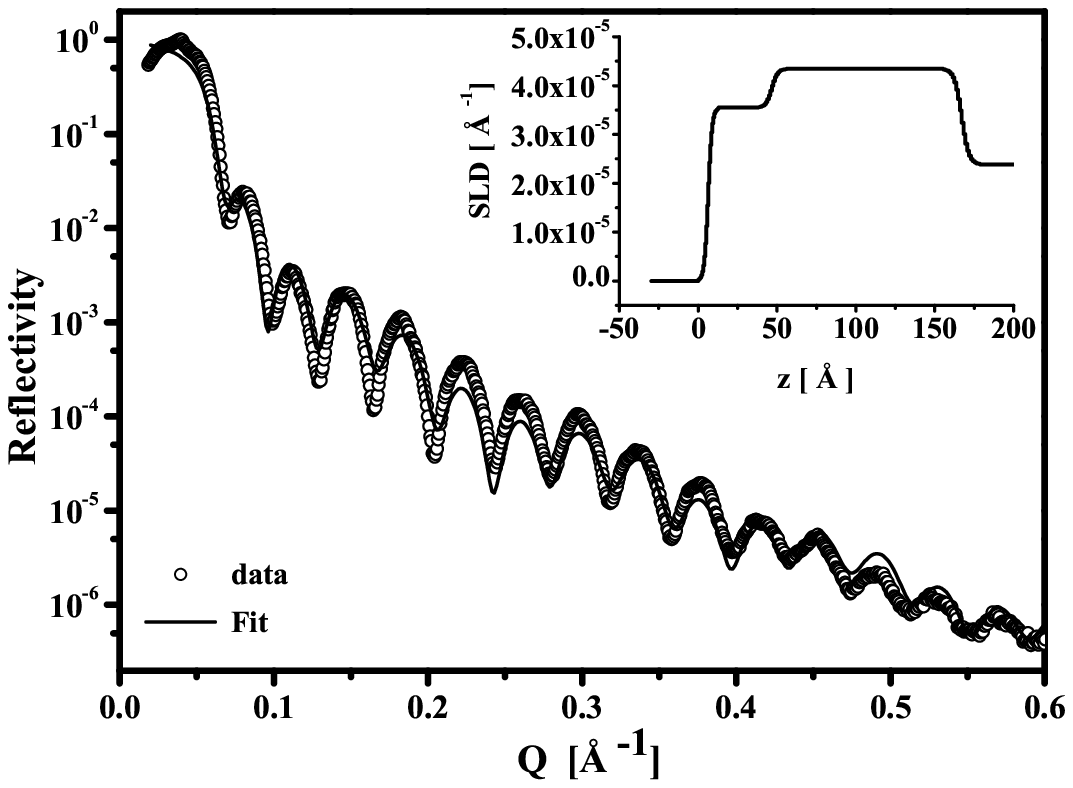}
\includegraphics[clip=true,keepaspectratio=true,width=1\linewidth]{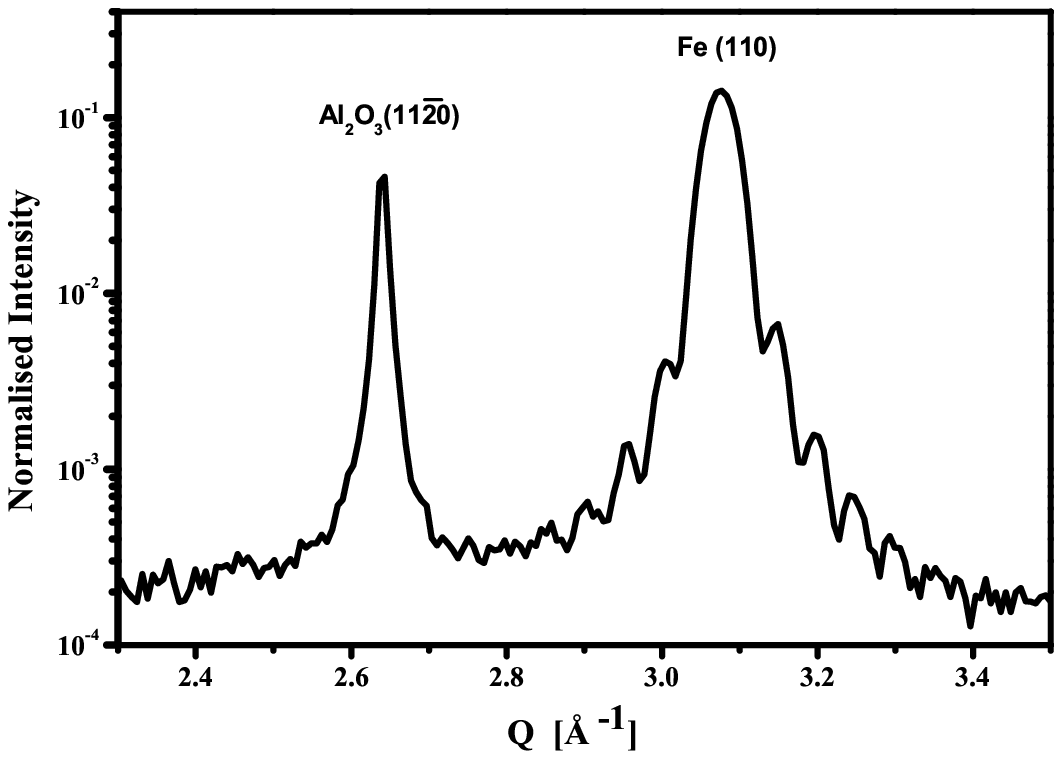}
\end{center}
\caption{\label{fig1} a) Reflectivity curve of the CoO/Fe bilayer
structure (symbols - the experimental data, line - the fit using a
model presented in the inset). b) The radial scan in the direction
normal to the sample surface around the
Al$_{2}$O$_{3}(11\bar{2}0)$ and Fe(110) Bragg peak positions. }
\end{figure}

Fig. \ref{fig1}a shows a reflectivity curve of the CoO/Fe bilayer
structure deposited  on a Al$_{2}$O$_{3}(11\bar{2}0)$ substrate.
The thickness oscillations corresponding to the Fe layer are
clearly visible up to $Q$ = 0.6 \AA$^{-1}$ with a small amplitude
modulations corresponding to the CoO layer, where $Q$ is the
scattering vector $Q = (4\pi/\lambda)sin\theta$. A fit of this
reflectivity curve using the Parratt formalism ~\cite{Parratt} and
the roughness model of Nev\'{o}t and Croce ~\cite{Nevot} gives the
electron density profile presented in the inset in
Fig.~\ref{fig1}a.

Fig. \ref{fig1}b shows the radial scan in the direction normal to
the diffraction planes, around the Al$_{2}$O$_{3}(11\bar{2}0)$ and
Fe(110) Bragg peak positions. The Laue oscillations on either
sides of the main Fe(110) Bragg peak are clearly seen. They reveal
both  a good crystalline quality of the  iron layer and an high
interface quality between the Fe and CoO layer. The smaller
intensity of Al$_{2}$O$_{3}(11\bar{2}0)$ substrate peak results
from a small and intentional misalignment. We have not found any
Bragg reflections from the CoO layer. Therefore we concluded that
this layer is polycrystalline. The deposition temperature of the
CoO was $200^{\circ}$~C, which is  the optimal growth temperature
for polycrystalline CoO, obtained from detailed growth studies
presented in Ref.~\cite{gnowak:2005}. The crystalline quality of
cobalt oxide was intentionally sacrificed in order to achieve a
smooth CoO/Fe interface, and to eliminate orientational
difficulties, characteristic for mono-crystalline CoO layers.

\section{Soft x-ray resonant magnetic scattering studies}

X-ray resonant magnetic scattering (XRMS) provides direct
information on the magnetic structure of materials. In order to
study XRMS of 3d transition metals, the \textit{L} absorption
edges must be utilized, which are located in the soft x-ray range.
Soft x-ray resonant magnetic scattering using either circularly or
linearly polarized x-rays, has proven to be a highly useful
technique for the study of magnetic properties of buried layers,
interfaces and, depth-dependent magnetic properties. Moreover, by
varying the external magnetic field applied parallel to the sample
plane and parallel to the x-ray circular helicity, close to
energies corresponding to $L$ absorption edges element-specific
hysteresis loops can be measured \cite{Kortright}. The XRMS
experiments were carried out at the undulator beamlines UE56/1\&2
and bending magnet beamline PM3 of BESSY. Since for this energy
range special vacuum conditions are required, a UHV-diffractometer
ALICE \cite{alice} was used for the experiments to be described
below. The magnetic field was applied in the scattering plane and
parallel to the sample surface.

\subsection{Reflectivity and asymmetry ratio}

\begin{figure}[!th]
\begin{center}
\includegraphics[clip=true,keepaspectratio=true,width=1\linewidth]{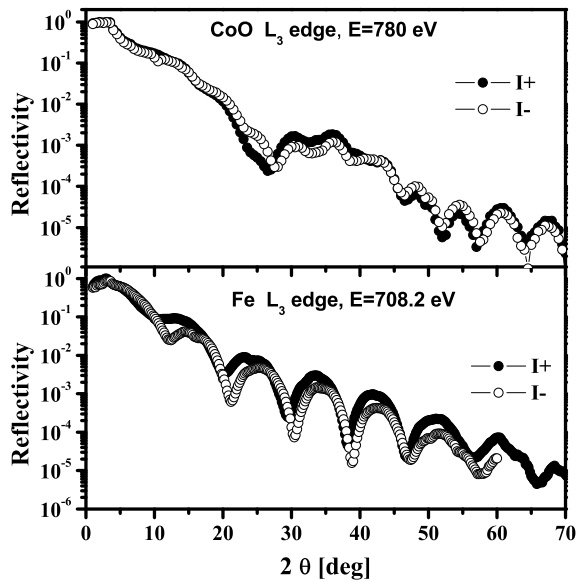}
\end{center}
\caption{\label{fig2} The reflectivities of CoO/Fe bilayer
measured at room temperature for the magnetic field applied in the
sample plane parallel (I$^{+}$, closed symbols) and antiparallel
(I$^{-}$, open symbols) to the photon helicity. The energy of the
circular polarized light was tuned close to the $L_{3}$ absorption
edges of Co (a) and Fe (b), respectively.}
\end{figure}

\begin{figure}[!th]
\begin{center}
\includegraphics[clip=true,keepaspectratio=true,width=1\linewidth]{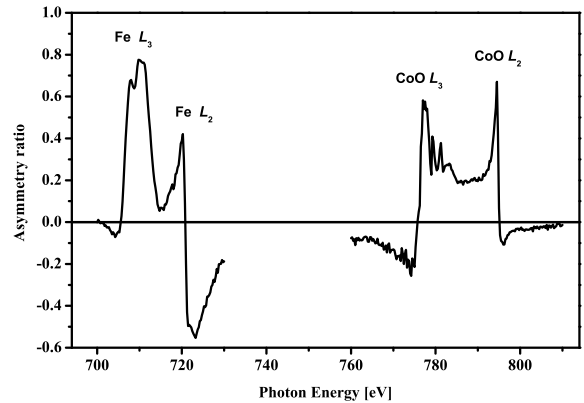}
\end{center}
\caption{\label{fig3} The energy dependence of the asymmetry ratio
for a fixed scattering angle of 2$\theta=~32^{\circ}$.}
\end{figure}

Fig. \ref{fig2} shows the reflectivities of CoO/Fe bilayer
measured at room temperature for the magnetic field applied in the
sample plane parallel (I$^{+}$) and antiparallel (I$^{-}$) to the
photon helicity. The photon energy of the circular polarized light
was tuned close to the $L_{3}$ absorption edge of Co
(Fig.~\ref{fig2}a) and Fe (Fig.~\ref{fig2}b), respectively. The
highest magnetic sensitivity is reached at the maxima of the
thickness oscillations. The energy dependence of the asymmetry
ratio A=(I$^{+}$-I$^{-}$)/(I$^{+}$+I$^{-}$) is depicted in
Fig.~\ref{fig3} for a fixed scattering angle of
2$\theta=~32^{\circ}$. Strikingly, the asymmetry of Co does not
vanishes as expected for AF materials. Moreover, even above the
N\'eel temperature (T$_N$=291~K) a ferromagnetic signal belonging
to the CoO layer is still visible.

The reflectivities  shown in Fig. \ref{fig2} reveal the sign of
the interfacial coupling of the two ferromagnetic materials in
question, which share a common interface or which are exchange
coupled over a nonmagnetic layer. For both layers the minima in
the I$^{+}$ reflectivity curve always lies at lower angles than
the minima in the I$^{-}$ curves.  This allows to clearly conclude
that the magnetization of both layers is parallel. An antiparallel
alignment of the layer would produce a reverse position of the
minima in the reflectivity curves. It should be pointed out that
element-specific anti-phase  hysteresis loops is not
characteristic for the coupling sign~\cite{hanke:2001}. This can
be seen from the reflectivity curves. At some incident angles the
asymmetry is positive for Fe and negative for CoO. Therefore,
taken the hysteresis loops alone they would suggest an AF-coupling
between the layers, whereas the true orientation is ferromagnetic,
as clearly seen in the reflectivity curves.

\subsection{Temperature dependence of the element-specific
hysteresis loops}

\begin{figure}[!th]
\begin{center}
\includegraphics[clip=true,keepaspectratio=true,width=1\linewidth]{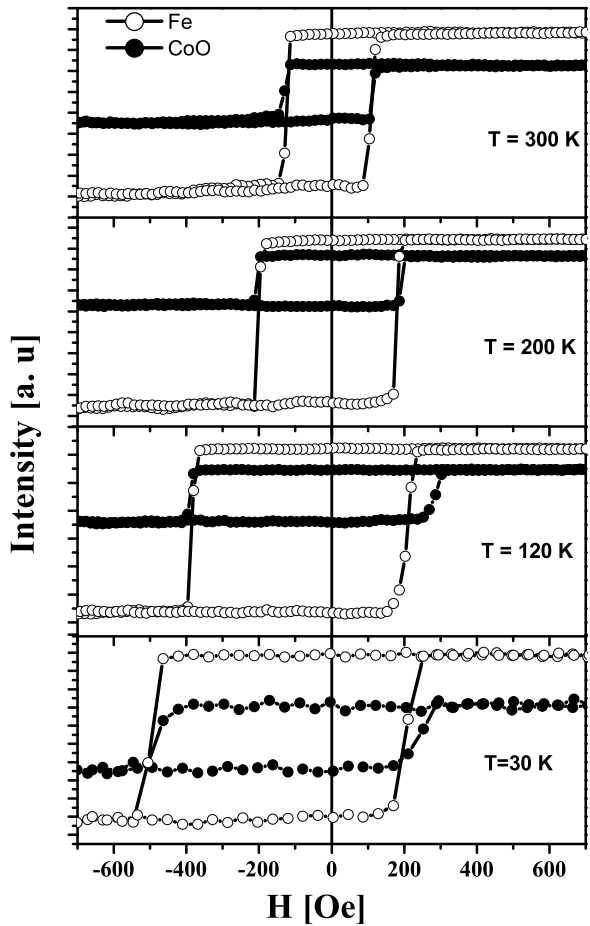}
\end{center}
\caption{\label{fig4} The temperature dependence of the  exchange
bias hysteresis loops measured at $L_{3}$ absorption edges of Co
(E=780 eV, closed symbols) and Fe (E=708.2 eV, open symbols).
Scattering angle is 2$\theta=~32^{\circ}$. }
\end{figure}

The exchange bias hysteresis loops measured at the $L_{3}$
absorption edges of Co (E=780 eV, closed symbols) and Fe (E=708.2
eV, open symbols) and for different temperatures are shown in
Fig.~\ref{fig4}. The measuring procedure is as follows: first the
system was heated up to T=300 K,~which is well above the N\'eel
temperature of CoO. Here, a field of +2000~Oe was applied parallel
to the sample plane and parallel to the helicity of the circular
polarization of the x-ray beam. Subsequently, the system was field
cooled to the lowest available temperature, which is about 30~K.
Here, several hysteresis loops were measured (not shown) in order
to eliminate training effects of the hysteresis loops. After
acquiring a stable reversible magnetization curve at 30~K, we
raised the temperature stepwise, from low to high T, and for each
temperature we measured one element-specific hysteresis loop at
the energies corresponding to Fe and CoO, respectively. The
hysteresis loops of Fe as a function of temperature show a typical
behavior. At low temperatures an increased coercive field and a
shift of the hysteresis loop is observed. As the temperature is
increased, the coercive field and the exchange bias decrease until
the blocking temperature is reached. Here, the exchange bias
vanishes and the coercive field shows little changes as the
temperature is further increased.
Strikingly, a ferromagnetic hysteresis loop corresponding to the
CoO layer is observed for all temperatures, following closely the
hysteresis loop of Fe (or vice versa), with some notable
differences. It appears that the ferromagnetic components of CoO
develop different coercive fields than Fe below the blocking
temperature. Moreover, the ferromagnetic moment of CoO is present
also above the N\'eel temperature. Here, the AF layer is in a
paramagnetic state, therefore the magnetic moment per spin is very
low for applied fields used here. It implies, that a ferromagnetic
layer should be present at the interface between Fe and CoO with a
Curie temperature, higher then the N\'eel temperature of CoO.

\subsection{XRMS comparison between CoO single layer and CoO/Fe bilayer system}

In order to exclude the presence of Co clusters in the CoO layer
as a possible explanations of our results presented above, we have
prepared two additional samples with the same sputtering
technique: one was the same CoO/Fe exchange bias system and
another one grown in the same run was just a simple CoO layer
deposited on a sapphire substrate. First both samples were field
cooled and then we measured at low temperatures the element
specific hysteresis loops at the $L_{3}$ absorption edge of Co.
For the CoO/Fe bilayer the previous results were reproduced. But
for the single CoO  layer no ferromagnetic signal was observed. We
infer from this the absence of cobalt clusters in the CoO
layer~\cite{radu-apl}. Vice versa, the ferromagnetic moment in the
CoO layer must be induced by the proximity between the F and the
AF layer. The induced magnetic moment appears to consist of two
parts. One part of the spins are strongly coupled to the AF
itself. They do not rotate with the rotation of the F layer.
Another part of these spins have a reduced AF anisotropy and due
to exchange coupling to the F layer, they show a similar
hysteresis loop as the F layer.

\section{Discussion}

The M\&B model~\cite{bean:1956} assumes that the AF spins rigidly
form an AF state, but they may slightly rotate as a whole during
the magnetization reversal of the F layer. Within the M\&B model,
enhanced coercivity is not accounted for. The interface is assumed
to be perfectly uncompensated with the interface AF spins having
the same anisotropy as the bulk spins. However, the interface is
never perfect. Roughness, deviations from stoichiometry,
interdiffusion, structural defects, etc. cause non-ideal magnetic
interfaces. It is therefore naturally to assume that,
statistically, a fraction of the AF spins have lower anisotropy as
compared to the bulk ones. These interfacial AF spins can rotate
together with the ferromagnet. They mediate the exchange coupling,
induce an enhanced coercivity, but soften the extreme coupling
condition assumed by M\&B. Therefore, we assume that the
anisotropy of the AF interface layer varies  from $K_{int}=0$ next
to the F layer to $K_{int}=K_{AF}$ next to the AF layer, where
$K_{AF}$ is the anisotropy constant of a presumably uniaxial
antiferromagnet. This variation of the anisotropy constant across
the interface  governs the enhanced anisotropy of the
ferromagnetic layer. So far it was  believed that the enhanced
coercivity in F/AF exchange biased systems is caused by
compensated AF spins at the F/AF interface. We argue that for most
of the AF materials a compensated or uncompensated spin having the
same anisotropy as the bulk AF layer would be practically
impossible to reverse by rotating the F layer. Therefore we need
to assume low anisotropy AF spins in order to qualitatively
describe the rotating AF seen in the  experimental data. Evidence
for the existence of low anisotropy AF spins at the F/AF interface
is provided here through measurements of element specific
hysteresis loops.

\section{Conclusions}

We have investigated the ferromagnetic behavior of the AF spins
for an Fe/CoO exchange bias bilayer. For all temperatures we
observe a non-vanishing ferromagnetic hysteresis loop of the spins
belonging to the AF layer. They reverse, in phase, with the spins
of the Fe layer, but display a different coercive field. We assume
that those AF spins are located at the interface between the F/AF
layer, and that they have, on average, a reduced AF anisotropy.
This assumption may lead to a better understanding of the enhanced
coercivity observed in almost all F/AF systems.

 We would like to
thank S. Erdt-B\"{o}hm for technical assistance during sample
preparation, Th. Kachel, B. Zada, W.Mahler (BESSY) and O. Seeck
(HASYLAB) for their help with the beamline operation. We
gratefully acknowledge support through the SFB 491 "Magnetische
Heteroschichten: Struktur und elektronischer Transport" of the DFG
and the German Federal Ministry of Education and Research (BMBF)
under Contracts No. 03ZA6BC2 (ALICE diffractometer) and No.
05ES3XBA/5 (support of travel to BESSY).

\end{document}